\PassOptionsToPackage{prologue,dvipsnames}{xcolor}
\documentclass[prologue,dvipsnames,aps,prl,twocolumn,showpacs,superscriptaddress,groupedaddress]{revtex4-1}
\usepackage{graphicx}
\usepackage{physics}
\usepackage{amssymb}
\usepackage{amsmath}
\usepackage{latexsym}
\usepackage{ulem}
\usepackage{color}
\usepackage{xcolor}
\usepackage{dcolumn}
\usepackage{subfigure}
\usepackage[T1]{fontenc}
\usepackage[utf8]{inputenc}
\usepackage[english,french]{babel}
\usepackage[colorlinks=true,linkcolor=blue,citecolor=blue]{hyperref} %
\usepackage{bm}        
\usepackage{amssymb}   

\usepackage[capitalize]{cleveref}

\usepackage{lipsum}

\begin{document}

\graphicspath{{figures/}}

\definecolor{airforceblue}{rgb}{0.36, 0.54, 0.66}

\title{Robustness of flat band superconductivity against disorder in a two-dimensional Lieb lattice model}\author{G. Bouzerar}%
\email{georges.bouzerar@neel.cnrs.fr}
\author{M. Thumin}%
 \email{maxime.thumin@neel.cnrs.fr}
\affiliation{Université Grenoble Alpes, CNRS, Institut NEEL, F-38042 Grenoble, France}%
\date{\today}

\begin{abstract}
Recently, the possibility of high-temperature superconductivity (SC) in flat-band (FB) systems has been the focus of a great deal of activity. 
This study reveals that unlike conventional intra-band SC for which disorder has a dramatic impact, that associated with FBs is surprisingly robust to disorder-induced fluctuations and quasi-particle localization. 
In particular, for weak off-diagonal disorder, the critical temperature decreases linearly with disorder amplitude for conventional SC, whereas it is only quadratic in the case of SC in FBs.
Our findings could have a major impact on the research and development of new compounds whose high purity will no longer be a critical barrier to their synthesis.
\end{abstract} 

\maketitle


\textit{Introduction.-} 
The last decade has seen the emergence of a new field of research in condensed matter physics where topology and quantum nature of matter intertwine: the physics of non-dispersive bands \cite{Leykam_review, Regnault2022, Neves_catalogue}. Flat-bands (FBs) may arise in special lattices such as the Lieb, kagome, pyrochlore, dice ... \cite{Lieb_photonic, Lieb_optical_lattice,Kagome_CoSn,pyrochlore,Dice_proposal}. They are naturally present in bipartite lattices with a different number of atoms/orbitals in the two sublattices and where hoppings obey specific conditions. Experimentally, FBs can be realized in Moiré materials such as van der Waals heterostructures by twisting one layer with respect to the other \cite{Cao2018_1, Cao_t3g, Efetov, tWse2_Dean}. FBs may appear as well in phononic, magnonic, metamaterials, or in cold atom systems \cite{magnonic_FB, phononic_metamaterial, Lieb_Goldman}. In FBs, the quenching of the kinetic energy leaves the electron-electron interaction as the only relevant energy scale, giving rise to strongly correlated quantum phases.
One of these is the surprising FB superconductivity (SC) which could make room-temperature SC an accessible reality at last. Because of the unconventional geometric nature of the SC in FB systems, the critical temperature ($T_c$) is expected to scale linearly with the effective attractive electron-electron interaction $T_c \propto |U|$ \cite{BCS_FB,Volovik_T_linear, Volovik_T_linear_Flat_bands_in_topological_media}, promising higher $T_c$'s than in standard BCS superconductors where $T_c^{BCS} \propto t \, e^{-1/\rho(E_F)|U|}$ \cite{BCS}. We emphasize, that these predictions rely essentially on mean-field approaches. However, numerous studies of an exact nature, such as the exact diagonalization (ED) \cite{Peotta_Lieb, Torma_Band_geometry}, the Density Matrix Renormalization group (DMRG) and the quantum Monte Carlo (QMC) approach \cite{Hofmann_PRB,Peri_PRL,Arbeitman_PRL}, have confirmed these predictions.
Although this is not an exact method, excellent agreement was observed between dynamical mean field theory (DMFT) and the standard mean field approach \cite{Peotta_Lieb, Torma_Band_geometry}.
Moreover, in one-dimensional systems, where quantum fluctuations are most important, the quantitative agreements between Bogoliubov de Gennes (BdG) and DMRG for the pairings and the superfluid weight (SFW) were truly impressive \cite{Batrouni_sawtooth, Batrouni_Designer_Flat_Bands}.  
Remark as well that exact results that support the use of mean field approach have been established in Refs~\cite{arbeitman_Letter, Peotta_SU(2)}
\\ These recent studies on FB superconductivity do not consider the impact of the disorder which cannot be ignored in real compounds. In conventional superconductors, it is well known that when a moderate amount of disorder is introduced the quasi-particle localization leads to a rapid suppression of the SFW and of the $T_c$ \cite{Maekawa_1982, Belitz_1987_McMillan, Belitz_1987, Belitz_1992, Finkelstein_1994, Ghosal_1998, Ghosal_2000, Antonenko_2020}.
\begin{figure}[h!]
\centering
\includegraphics[scale=0.5]{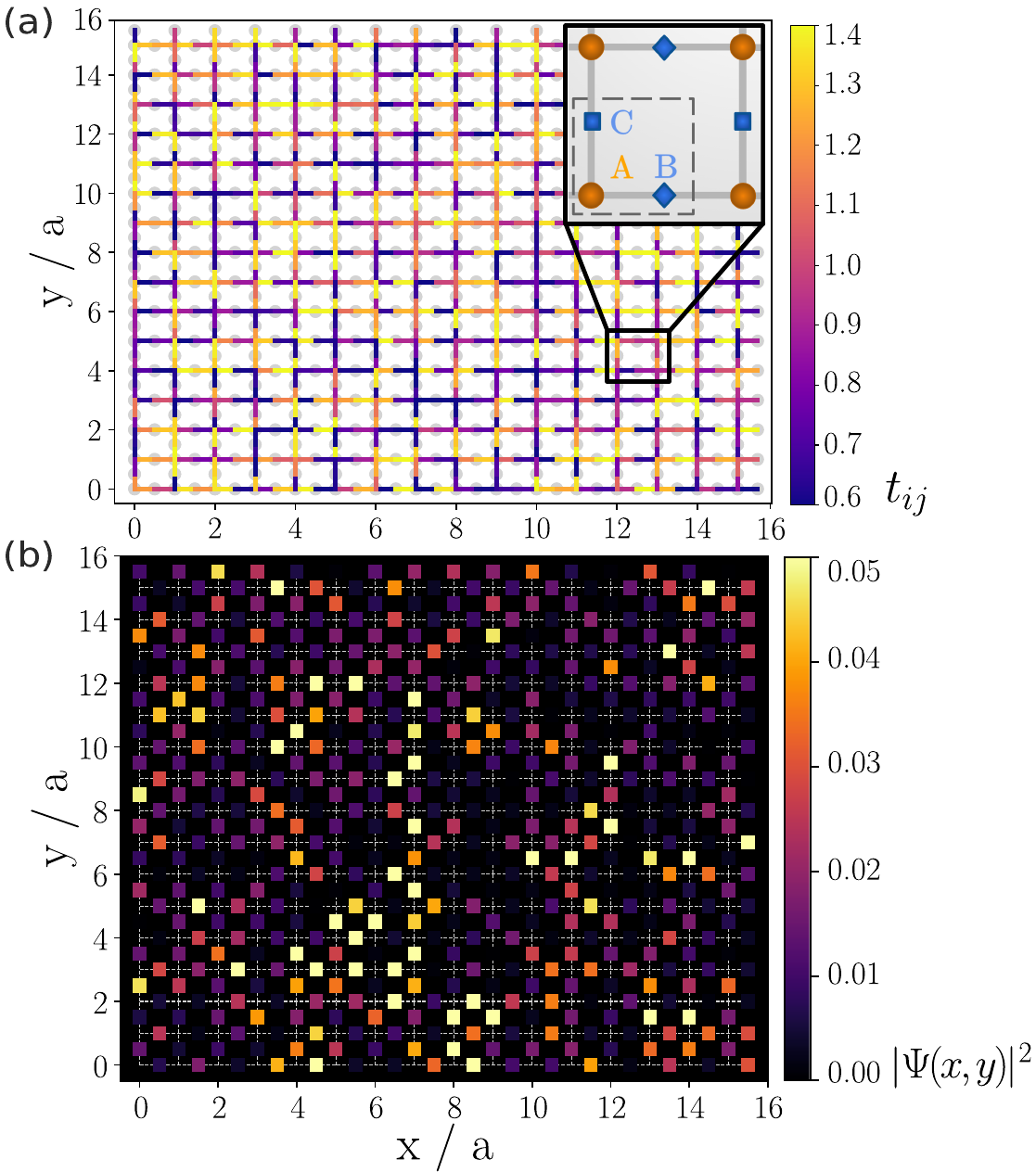} 
\caption{ \textbf{(a)} Lieb lattice for a given configuration of disorder ($W/t=0.8$). Grey symbols are the orbital positions, colored bonds give the hopping values in the half-filled lattice. \textbf{(b)} Map of the square of the wave-function amplitude of a typical FB state. Only $\mathcal{B}$ sublattice is represented.
}
\label{Fig. 1}
\end{figure}
\begin{figure*}
\centering
\includegraphics[scale=0.4]{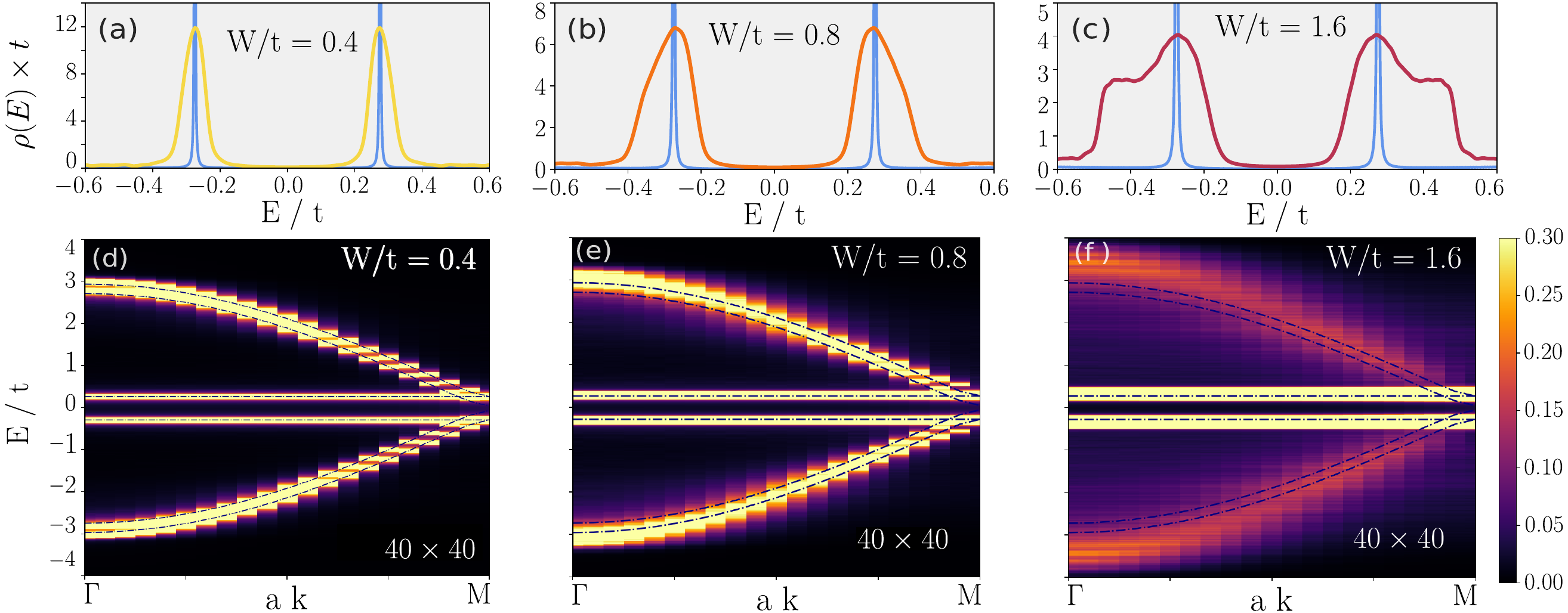}    
    \caption{ \textbf{(a-c)} Quasi-particle density of states for $W/t=0.4, 0.8$ and $1.6$. The blue continuous lines correspond to $W/t=0$.
    \textbf{(d-f)} Spectral functions along the $\Gamma-M$ direction for the same values of the disorder strength as in panels (a-c). The system contains $40\times40$ unit cells. The dashed blue lines are the dispersions for $W/t=0$. } 
\label{Fig. 2}
\end{figure*} 
Recently, in BCS systems, it has been shown that the suppression of $T_c$ scales as $\frac{\Delta T_c}{T^{0}_c} \propto -\frac{1}{k_Fl_e}$ where $T^{0}_c$ is that of the clean system, $k_F$ is the Fermi wavevector and $l_e$ the mean free path \cite{Antonenko_2020}. Whereas conventional SC is of intra-band type, that in FBs is governed by inter-band processes. Thus, it is not obvious whether the suppression of the SC phase by disorder will be stronger or not when the Fermi level is located in the FBs. In this work, our purpose is to address precisely this issue. The prototype chosen here is the bipartite Lieb lattice which corresponds to that of the CuO$_2$ planes in cuprates. It consists in 3 orbitals per unit cell $A$, $B$ and $C$ as illustrated in Fig.$\,$\ref{Fig. 1}(a). We define $\mathcal{A}$ the first sublattice made up of $A$-type orbitals while the second $\mathcal{B}$ includes $B$- and $C$-type orbitals.
Because of Hilbert space size limitation, and high CPU cost, it is impossible to treat the disorder exactly in a real-space approach, within the framework of exact methods. However, since the BdG approach has proved reliable and accurate in clean systems, we choose to tackle our issue within the Real-Space-BdG (RS-BdG).
This allows a complete treatment of the localization effects, the access to large systems, a systematic average over disorder realizations, and the evaluation of the size effects. It should be noticed that FB superconductivity in the clean Lieb lattice have been addressed in Refs$\,$\cite{Batrouni_CuO2,Peotta_Lieb}.

\textit{Model and methods.-}
Electrons are described by the attractive Hubbard model (AHM),
\begin{equation}
\begin{split}
    \hat{H} = \hspace{-0.2cm}\sum_{\langle i\lambda,j\eta \rangle, \sigma} \hspace{-0.3cm} t^{\lambda\eta}_{ij, \sigma} \; \hat{c}_{i\lambda, \sigma}^{\dagger} \hat{c}_{j \eta, \sigma} - \mu \hat{N} - |U| \sum_{i\lambda} \hat{n}_{i\lambda\uparrow}\hat{n}_{i\lambda\downarrow}
    \label{H_exact}
\end{split}
,
\end{equation}
where $\hat{c}_{i \lambda \sigma}^{\dagger}$ creates an electron of spin $\sigma$ at site $\textbf{r}_{i\lambda}$, $i$ being the cell index and $\lambda=A,B$ and $C$ the orbitals. Hopping integrals are non-zero for nearest-neighbor pairs $\langle i\lambda,j \eta\rangle$ only and  $t^{\lambda\eta}_{ij} = -t + \frac{W}{2} \cos{(\theta_i^\lambda - \theta_j^\eta)}$ where $W$ measures the disorder strength and $\theta_i^\lambda$ are uncorrelated random variables chosen in the interval $[-\pi,\pi]$. Finally, $\hat{N}=\sum_{i\lambda,\sigma} \hat{n}_{i\lambda,\sigma}$ is the particle number operator, $\mu$ is the chemical potential, and $|U|$ is the strength of the on-site attractive electron-electron interaction. For what follows, it is convenient to define three regimes: (i) weak disorder $W/t \le 0.5$ (ii) intermediate disorder $0.5 \le W/t \le 2$ (iii) strong disorder $W/t \ge 2$. In this study, we are only interested in the impact of the disorder, hence we set $|U| = t = 1$.
\\
Before turning to our results, we briefly discuss our model of disorder. In minimal model approaches, one often uses the Anderson model (AM) which consists in introducing uncorrelated random on-site potentials. Here, it would simply destroy the FB eigenstates and thus the FB physics.
Furthermore, in numerous situation, the AM seems extreme and even inappropriate for a realistic modeling. 
Indeed, if defects are point-like, such as vacancies, compensation defects, local dislocations and substituting atoms, a binary model (BM) of the disorder would be more suitable. In addition, contrary to the AM the FB physics
would not be washed away. As an illustration, in III-V diluted magnetic semiconductors, only the BM could capture the essential physics: presence of resonant states, ferromagnetic nature of the magnetic couplings, high Curie temperatures and transport properties as well Ref.$\,$\cite{Bouzerar_2007, Bouzerar_2010, Bouzerar_2016}. These features would be absent if the AM is used instead. In this work, we choose to introduce the disorder in the hoppings in order to preserve the bipartite character of the lattice \cite{Lieb_2_theorem}, which could mimic a system where the orbitals are misaligned.

We briefly recall that the RS-BdG approach consists in decoupling in the AHM the interaction term as follows,
\begin{equation}
\begin{split}
\hat{n}_{i\lambda,\uparrow}\hat{n}_{i\lambda,\downarrow} \overset{BdG}{\simeq}
&\langle\hat{n}_{i\lambda,\downarrow}\rangle  \hat{n}_{i\lambda,\uparrow} +\langle\hat{n}_{i\lambda,\uparrow}\rangle  \hat{n}_{i\lambda,\downarrow} \\
- \; &\,\frac{\Delta_{i\lambda}}{|U|} \hat{c}^{\dagger}_{i\lambda,\uparrow}\hat{c}^{\dagger}_{i\lambda,\downarrow} 
- \frac{\Delta_{i\lambda}^*}{|U|} \hat{c}_{i\lambda,\downarrow}\hat{c}_{i\lambda,\uparrow} + C_{i\lambda} \\ 
\end{split}
\end{equation}
where $C_{i\lambda}= - \langle\hat{n}_{i\lambda,\uparrow}\rangle  \langle \hat{n}_{i\lambda,\downarrow} \rangle  - \Big|\frac{\Delta_{i\lambda}}{U}\Big|^2$.
The $6 N_c$ self-consistent parameters $\langle \hat{n}_{i\lambda,\sigma} \rangle $ and $\Delta_{i\lambda}=-|U| \langle \hat{c}_{i\lambda,\downarrow} \hat{c}_{i\lambda,\uparrow} \rangle$ with $i=1,2,...N_c$ and $\lambda=A,B$ and C are respectively the local occupations and pairings, $N_c$ being the number of unit cells. $\langle\ldots\rangle $ corresponds to the grand canonical average at finite temperature. We consider here a non magnetic ground-state, which implies that $\langle\hat{n}_{i\lambda,\uparrow}\rangle=\langle\hat{n}_{i\lambda,\downarrow}\rangle=\langle\hat{n}_{i\lambda}\rangle/2$. The filling factor is defined as $\nu=N_e/N_c$ where $N_e$ is the number of electrons. The half-filled case corresponds to $\nu=3$. \\
From a technical point of view, for a fixed temperature and chosen configuration of the disorder (set of random hoppings), the determination of the self-consistent parameters requires the multiple full diagonalization of large complex matrices of size $6N_c\times6N_c$. In the second step, the observables (local pairings, local occupations and SFW) are averaged over many configurations of the disorder. Here, $N_c$ ranges from $16\times16$ to $40 \times 40$. Notice that this type of calculations is very demanding in terms of CPU since the number of diagonalizations required for a fixed temperature and given carrier density scales as $(6N_c)^3\times N_{iter}\times N_{dis}$ where $N_{iter}$ is the number of iterations before the convergence of the $6N_c$ self-consistent parameters is reached
and $N_{dis}$ being the number of configurations of disorder. Furthermore, $N_{iter}$ increases rapidly as the temperature increases.

\textit{Spectral function and density of states.-}
Figure$\,$\ref{Fig. 1}(a) depicts a configuration of the hoppings for a moderate amplitude of the disorder $W/t=0.8$. For this specific configuration,
figure$\,$\ref{Fig. 1}(b) illustrates the typical structure of a FB quasi-particle (QP) eigenstate. Only $B$ and $C$ 
orbitals are displayed, on $A$ orbitals the weight being exactly zero (bipartite lattice). For this value of $W/t$ the QP eigenstate is rather inhomogeneous, but at the same time it is relatively widespread over the entire lattice. In Fig.$\,$\ref{Fig. 2}(a-c) the density of states in the vicinity of the QP gap is depicted for several values of $W$. An average over about $10$ configurations of disorder has been realized. We observe that as $W$ increases the width of the QP FB peak broadens rapidly and even, for the largest value of $W$, it exhibits a well defined shoulder.
\\
To complete our study we have plotted the spectral function (SF) $A(\textbf{k},\omega)$ in Fig.$\,$\ref{Fig. 2}(b),(c) and (d) for the same values of $W$ where $A(\textbf{k},\omega)=-\frac{1}{\pi} \langle \Im(G(\textbf{k},\omega))\rangle_{dis}$, $G(\textbf{k},\omega)$ being the Fourier transform of the disordered one-particle Green's functions and $\langle ... \rangle_{dis}$ means average over disorder.
First, we discuss the case of the dispersive QP bands. For weak disorder, away from the $\Gamma$-point, the QP modes are well defined with a small broadening. As $W$ increases, Fig.$\,$\ref{Fig. 2}(e) shows that the peaks widens even further, the region where the QPs are well defined shrinks significantly. Finally, when $W/t=1.6$, for a given $\textbf{k}$, one observes a single broad QP peak in the SF. We now consider the quasi FB eigenstates. We recall that for $W=0$ the QP FBs are well defined with energy is $\pm \Delta_\mathcal{B}$ (pairing on $\mathcal{B}$-lattice). When $W$ is switched on, the SF reveals that the concept of FB is still meaningful in the disordered lattice. Indeed, if we restrict ourself to the FB region, then the first moment $M_1(\textbf{k})= \int \omega \Tilde{A} (\textbf{k},\omega) d\omega$ is $\textbf{k}$-independent where $\Tilde{A}(\textbf{k},\omega) = \frac{A(\textbf{k},\omega)}{\int A(\textbf{k},\omega) d\omega}$. In addition, the width $\Gamma_{fb}$ of the QP eigenstate (inverse of the lifetime) appears momentum independent: $\Gamma_{fb} (\textbf{k})=\sqrt{M_2(\textbf{k})-M^2_1(\textbf{k})}$ where $M_2(\textbf{k})=\int \omega^2 \Tilde{A}(\textbf{k},\omega) d\omega$. Numerically, we find $M_{1}(\textbf{k})/t = \pm 0.279, \pm 0.287$ and $\pm 0.317$, $\Gamma_{fb}(\textbf{k})/t = 0.05$, $0.06$ and $0.09$ for respectively $W/t = 0.4, 0.8$ and $1.6$.
 
\begin{figure}
    \centering
\includegraphics[scale=0.5]{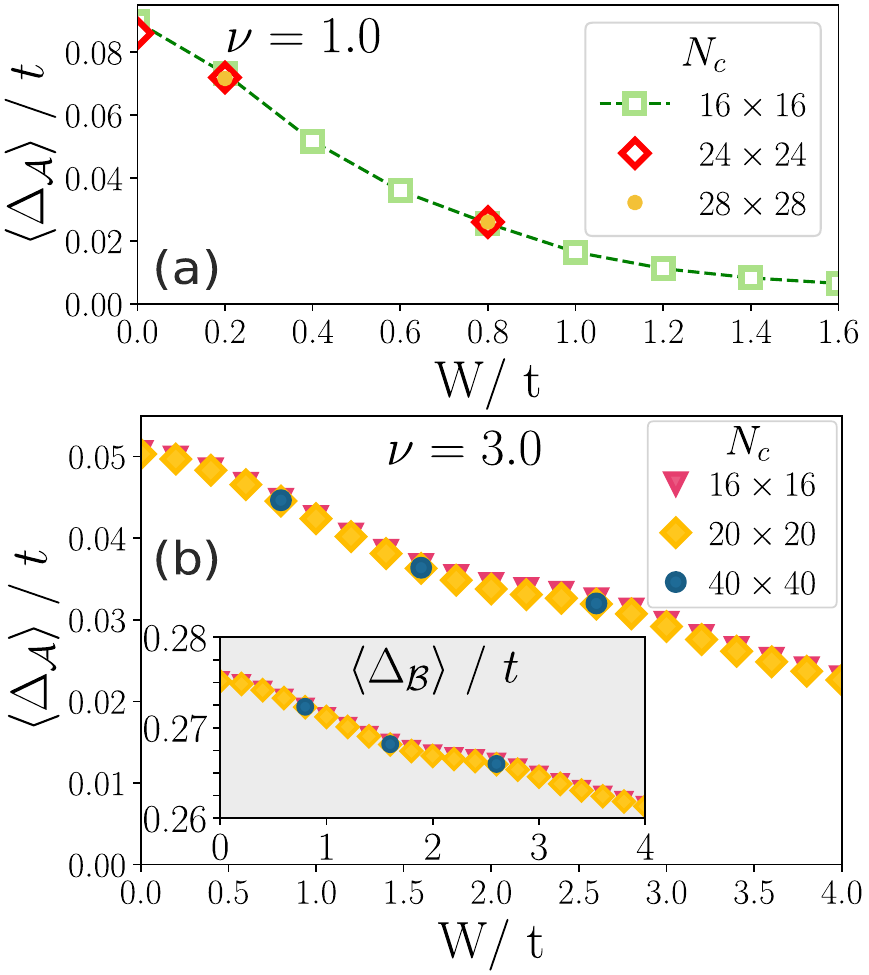}    
     \caption{Average pairing on sublattice $\mathcal{A}$
     for three different system sizes, for $\nu=1.0$ \textbf{(a)} and for $\nu=3.0$ \textbf{(b)}. The inset represents the average pairing on sublattice $\mathcal{B}$.
     }
 \label{Fig. 3}
\end{figure}
\textit{Pairings.-}
\begin{figure}[h!]
    \centering
    \includegraphics[scale=0.55]{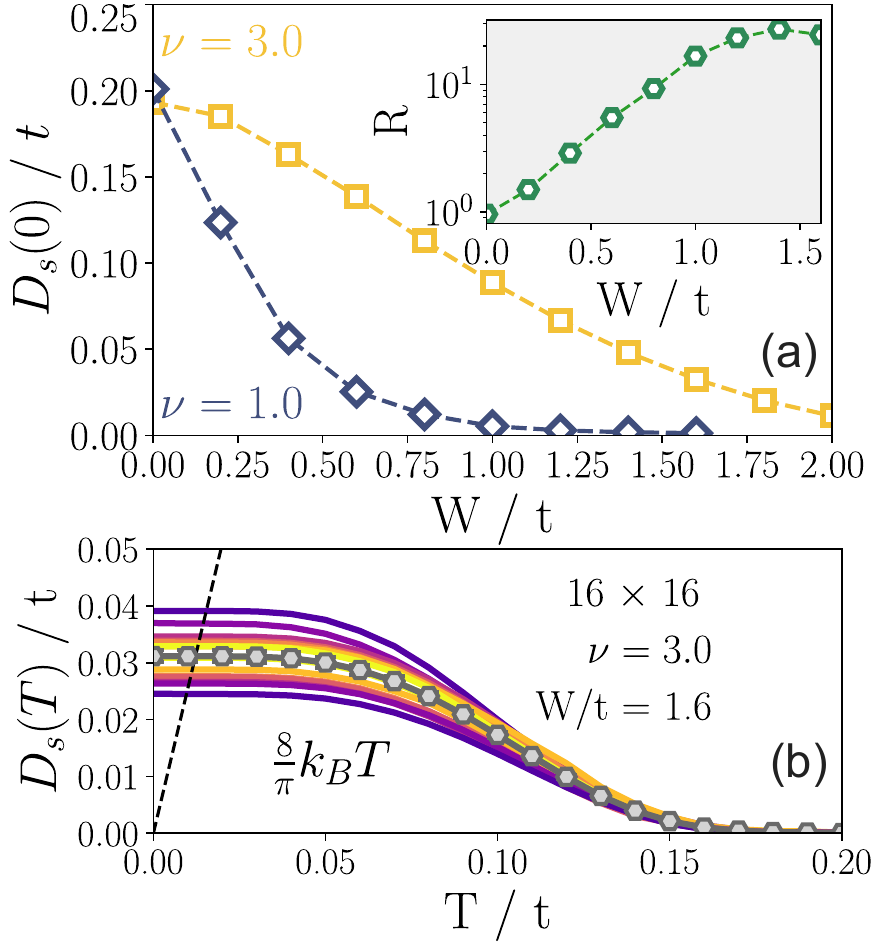}
    \caption{ \textbf{(a)} SFW at $T = 0$ as a function of $W/t$ at $\nu=1.0$ and $3.0$. The inset represents the ratio (R) of the SFW at $\nu=3.0$ divided by that obtained at $\nu=1.0$. \textbf{(b)} $D_s$ as a function of temperature for $\nu=3.0$, for $W/t=1.6$ and $N_c=16\times16$. Grey circles are the average values, the continuous lines correspond to $15$ different realizations of the disorder.}
    \label{Fig. 4}
\end{figure}
In this section we discuss the effect of disorder on the pairings. We consider two situations: (i) the Fermi energy is located inside the dispersive band ($\nu=1$), and (ii) the half-filled case ($\nu=3$). The first situation corresponds to the conventional intra-band SC and the other to the inter-band FB superconductivity.
We discuss the first scenario. In Fig.$\,$\ref{Fig. 3}(a) the average value of the pairing in $\mathcal{A}$-sublattice is plotted. As it can be seen, we find no finite size effects. In addition, $\langle \Delta_\mathcal{A} \rangle$ decays rapidly and monotonically as $W$ increases.
Indeed, it is found that $\langle \Delta_\mathcal{A} \rangle = 0.09t$ for $W/t=0$, while $\langle \Delta_\mathcal{A} \rangle = 0.015t$ for $W/t=1$, and hence the pairing has been strongly suppressed by about $83 \%$. A similar trend has been observed in Ref.$\,$\cite{Ghosal_2001}. We remark that $\langle \Delta_\mathcal{B} \rangle$ is not shown, since as found numerically in previous studies its average values obey the sum-rule \cite{Thumin_EPL_2023},
\begin{equation}
    N_\mathcal{B} \langle \Delta_\mathcal{B} \rangle =  N_\mathcal{A} \langle \Delta_\mathcal{A} \rangle,
    \label{eqfbcase}
\end{equation} 
where $N_\mathcal{A,B}$ being the number of orbitals in each sublattice, $N_\mathcal{A}/N_\mathcal{B} = 1/2$. We have checked that Eq.$\,$\ref{eqfbcase} is accurately verified for any realization of the disorder. We now consider the case of FB superconductivity as depicted in Fig.$\,$\ref{Fig. 3}(b). In contrast to the previous situation,
$\langle \Delta_\mathcal{A} \rangle$ decays very slowly as $W$ increases, even in the regime of strong disorder ($W/t \ge 2$). For $W/t=0$, the pairing $\langle \Delta_\mathcal{A} \rangle = 0.05 t$ while $\langle \Delta_\mathcal{A} \rangle = 0.043 t$ for $W/t=1$, it is reduced by $ 13\%$ only. On the other hand, for $\nu=3$, as analytically shown in Ref.$\,$\cite{Bouzerar_SciPost_2024} one finds,
\begin{eqnarray}
N_{\mathcal{B}} \langle \Delta_{\mathcal{B}} \rangle =  N_{\mathcal{A}} \langle \Delta_{\mathcal{A}}\rangle + \frac{|U|}{2} (N_{\mathcal{B}}-N_{\mathcal{A}}) .
\end{eqnarray} 
As a consequence, $\Delta_{\mathcal{B}}$ is strongly robust against the disorder, it varies very weakly even in the strong disorder regime, which contrasts with the conventional case.

\begin{figure}
    \centering
\includegraphics[scale=0.55]{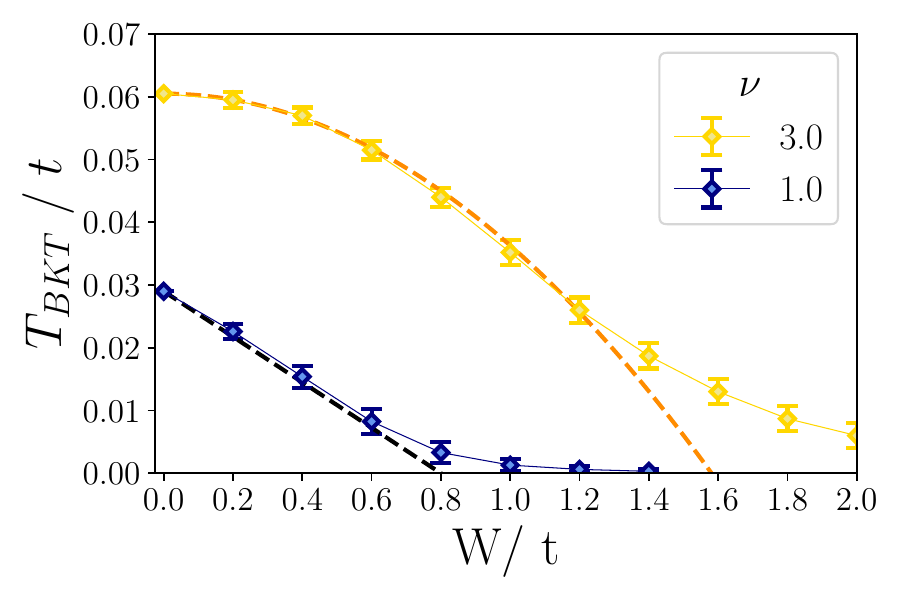}
     \caption{BKT temperature (symbols) as a function of $W/t$ for $\nu=1.0$ and $3.0$. Dashed lines are linear and quadratic fits: $y=0.06 \times \Big(1-0.40(W/t)^2 \Big)$ and $y=0.029 \times \Big(1-1.25(W/t) \Big)$. \\
     }
 \label{Fig. 5}
\end{figure}
\textit{Superfluid weight and Berezinskii–Kosterlitz–Thouless transition temperature.-}
In this section, our purpose is to discuss the impact of the disorder on the order parameter of the SC phase, the superfluid weight $D_s$ \cite{Khon_Ds, Shastry_Sutherland_Ds, Scalapino_Ds} which is carefully defined as \cite{Torma_revisiting},
\begin{equation}
D_s^{\mu\lambda}=\frac{1}{N_c}\frac{d^2\Omega (\textbf{q})}{dq_\mu dq_\lambda} \Big|_{|\textbf{q}|=0},
\end{equation}
$\Omega(q)$ is the grand-potential, $\textbf{q}$ is introduced by Peierls substitution. We emphasize that the derivative is total, it includes explicit and implicit (within the self-consistent parameters) $\textbf{q}$-dependences. Here, we consider $\mu=\lambda=x$ and simplify our notation $D_s^{xx}=D_s$.
In Fig.$\,$\ref{Fig. 4}(a), the average (over disorder) of the SFW at $T=0$ is plotted as a function of $W$ for $\nu=1$ and $\nu=3$.
First we observe that both decay as the disorder increases, but the reduction is extremely fast in the case of conventional SC. The slope for $W \rightarrow 0$ is $\frac{\partial D_s}{\partial W} \vert_0 = -0.4$. We can compare our results with those of Ref.$\,$\cite{Ghosal_2001} where  the calculations were performed for the square lattice. For a comparable filling factor and $|U|/t = 1.5$, it is found $\frac{\partial D_s}{\partial W} \vert_0 = -1.20$ which is significantly stronger compared to our finding. This could be explained by the fact that the disorder is diagonal in Ref.$\,$\cite{Ghosal_2001}. Fig.$\,$\ref{Fig. 4}(a) shows that the slope is much smaller in the case of half-filled FBs. The inset which depicts the ratio of the SFW at $\nu=3$ divided by that at $\nu=1$, allows to visualize how the disorder impact is more dramatic in the conventional SC case.
Figure$\,$\ref{Fig. 4}(b) depicts $D_s(T)$ as a function of $T$ for $\nu=3$ and $W/t = 1.6$. As it can be seen, $D_s(T)$ as a conventional shape. For this value of $W$, the Berezinsky-Kosterlitz-Thouless (BKT) transition temperature defined by the well known criterion $D_s(T_{BKT}) = \frac{8}{\pi} k_B T_{BKT}$ \cite{Jump_Ds,Peotta_Nature} is 
$T_{BKT} \approx 0.0125~t$. The fluctuations in the values of $T_{BKT}$ due to the disorder being here of the order of $30\%$.
\\
For a more quantitative assessment of the impact of the disorder, the average of $T_{BKT}$ is plotted as a function of $W$ for $\nu=1$ and $\nu=3$ in Fig.$\,$\ref{Fig. 5}. First, in the case of conventional intra-band superconductivity ($\nu=1$) we observe a rapid (linear) decay as the disorder increases. More precisely, for $W\le 0.8~t$, our data reveal that $r=(T_{BKT}(W)-T_{BKT}(0))/T_{BKT}(0)=-1.25\frac{W}{t}$. In contrast, in the FB case, the reduction is much weaker, it is only quadratic in $W$, more precisely we find $r =-0.40(\frac{W}{t})^2$.  Unambiguously, the FB superconductivity is systematically more robust in the face of disorder. 
By way of comparison, in the FB case, for $W/t=0.8$ the BKT temperature remains relatively high, at around 75\% of its value in the clean system. By contrast, at $\nu=1$, it has fallen sharply below $10\%$ of its value at $W=0$.
We remark that, in the case where the disorder corresponds to randomly distributed vacancies, we expect an effect similar to that found for off-diagonal disorder.

\begin{figure}
    \centering
\includegraphics[scale=0.53]{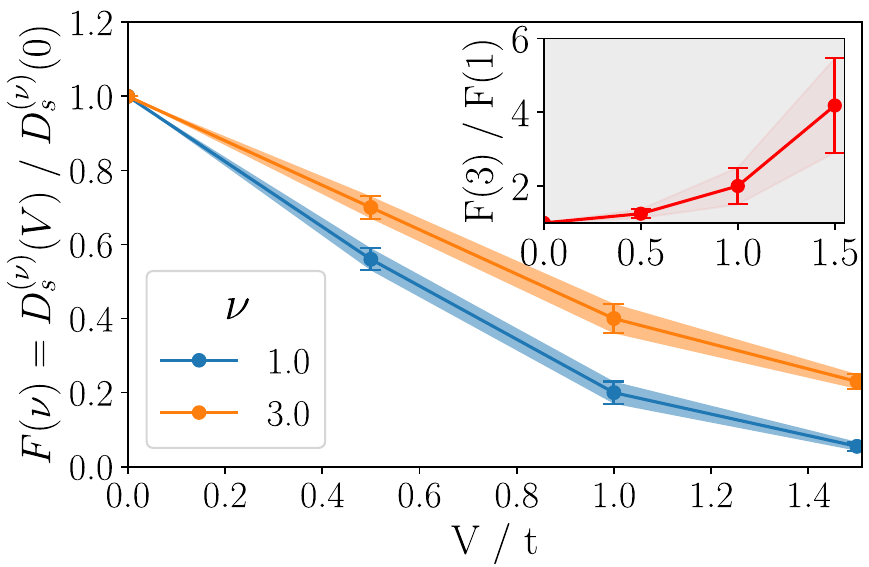}
     \caption{
     Superfluid weight rescaled by its value in the clean system ($F(\nu)$) as a function of the amplitude of the on-site disorder ($V/t$), for $\nu=1$ and $3$. The inset depicts the ratio $F(3)/F(1)$ for the same values of $V/t$.}
 \label{Fig. 6}
\end{figure}

\textit{Diagonal disorder.-}
We know that the introduction of diagonal disorder breaks the bipartite character of the lattice and destroys the FB eigenstates.
Therefore, it is interesting to see whether in such a case the superconductivity at $\nu=3$ is more or less robust to disorder than that at $\nu=1$. We now set $W=0$ and add to the Hamiltonian given in Eq.\eqref{H_exact} the following term,
\begin{eqnarray}
\hat{H}_{dis}= \sum_{i\lambda,\sigma}  \epsilon_{i\lambda}
\hat{c}_{i\lambda, \sigma}^{\dagger} \hat{c}_{i \lambda, \sigma} 
,
\end{eqnarray}
where the uncorrelated variables $\epsilon_{i\lambda}$ are randomly chosen in the interval $[-V/2,V/2]$, V measures the strength of the on-site disorder.\\
Fig.\eqref{Fig. 6} depicts the superfluid weight $D^{(\nu)}_s$ at $T=0~K$ as a function of $V$ for $\nu=1$ and $\nu=3$. As illustrated the effect of on-site disorder is clearly stronger in the case of conventional SC. More precisely the ratio $\frac{D^{(3)}_s(V)/D^{(3)}_s(0)}{D^{(1)}_s(V)/D^{(1)}_s(0)}$ increases rapidly as $V$ increases. Indeed, it is found that the ratio is approximately $2$ for $V/t=1$ while for $V/t=1.5$ it reaches $4$.
Thus, even in the extreme case of random on-site disorder, the superfluid weight remains much less sensitive to the effects of disorder than in the case where SC is essentially of conventional nature. 

\textit{Conclusion.-}
This work shows that the inter-band (or geometric) nature of FB superconductivity makes it much more robust against the effects of disorder than conventional (intra-band) superconductivity. 
More precisely, for off-diagonal disorder, it is revealed that the reduction of $T_c$ is linear with disorder amplitude in the case of conventional superconductivity , whereas in FBs it is only quadratic. We argue, that these findings could be promising and could have an important impact on both research and synthesis of new materials. Indeed, the most daunting challenge for the experts in materials growth and synthesis is to obtain the purest compounds, i.e. with the lowest possible concentration of native defects.


\end{document}